\documentclass{article}

\usepackage[letterpaper,top=2cm,bottom=2cm,left=3cm,right=3cm,marginparwidth=1.75cm]{geometry}

\usepackage{booktabs}
\usepackage{amsmath}
\usepackage{mathrsfs}
\usepackage{caption}
\usepackage{subcaption}
\usepackage{graphicx}
\usepackage{cite}

\title{High-degree Polynomial Noise Subtraction for Disconnected Loops}

\author{Paul Lashomb\footnotemark[2]
\and Ronald B. Morgan\footnotemark[3] 
\and Travis Whyte\footnotemark[4]
\and Walter Wilcox\footnotemark[5]
}

\begin{document}

\maketitle

\renewcommand{\thefootnote}{\fnsymbol{footnote}}
\footnotetext[1]{WW acknowledges support from the Baylor Summer Research Award program.}
\footnotetext[2]{Department of Physics, Baylor University, Waco, TX 76798 ({\tt Paul\_Lashomb@baylor.edu}).}
\footnotetext[3]{Department of Mathematics, Baylor University, Waco, TX 76798 ({\tt Ronald\_Morgan@baylor.edu}).}
\footnotetext[4]{Department of Applied Mathematics and Theoretical Physics, University of Cambridge, Cambridge, CB3 0WA, UK ({\tt tw601@cam.ac.uk}).}
\footnotetext[5]{Department of Physics, Baylor University, Waco, TX 76798 ({\tt Walter\_Wilcox@baylor.edu}).}
\renewcommand{\thefootnote}{\arabic{footnote}}

\begin{abstract}


In lattice QCD, the calculation of physical quantities from disconnected quark loop calculations have large variance due to the use of Monte Carlo methods for the estimation of the trace of the inverse lattice Dirac operator. In this work, we build upon our POLY and HFPOLY variance reduction methods by using high-degree polynomials. Previously, the GMRES polynomials used were only stable for low-degree polynomials, but through application of a new, stable form of the GMRES polynomial, we have achieved higher polynomial degrees than previously used. While the variance is not dependent on the trace correction term within the methods, the evaluation of this term will be necessary for forming the vacuum expectation value estimates. This requires computing the trace of high-degree polynomials, which can be evaluated stochastically through our new Multipolynomial Monte Carlo method. With these new high-degree noise subtraction polynomials, we obtained a variance reduction for the scalar operator of nearly an order of magnitude over that of no subtraction on a $24^3 \times 32$ quenched lattice at $\beta = 6.0$ and $\kappa = 0.1570 \approx \kappa_{crit}$. Additionally, we observe that for sufficiently high polynomial degrees, POLY and HFPOLY approach the same level of effectiveness. We also explore the viability of using double polynomials for variance reduction as a means of reducing the required orthogonalization and memory costs associated with forming high-degree GMRES polynomials.

\end{abstract}


\newpage 

\section{Introduction}

Signals obtained from lattice QCD are related to the trace of the inverted Wilson-Dirac operator. The large expense of inverting the Wilson-Dirac operator makes such a calculation impractical to perform exactly. For this reason, Monte Carlo methods are typically employed to estimate the trace stochastically. This Monte Carlo estimation of the trace of a matrix is commonly done through the use of Hutchinson's method \cite{hutchinson89}. This comes at the cost of introducing a large degree of noise into the signal for disconnected diagrams. One way to combat this is through noise subtraction methods \cite{thron98}. We have previously \cite{baral} developed and implemented several types of noise subtraction methods. The results for our polynomial subtraction (POLY), and our combination method which combines polynomial noise subtraction with hermitian-forced eigenvalue subtraction  (HFPOLY), however, were limited by the degree of the polynomial due to the instability of the method \cite{liupp} used to form them. In this work, we build on our previous POLY and HFPOLY methods by implementing a new, stable, high-degree polynomial from the GMRES polynomial. This method is discussed in further detail in Ref. \cite{embree21} and \cite{morgan19} in the context of polynomial preconditioning. In this work, we make use of such high-degree polynomials for the purpose of variance reduction of the trace estimator, and we observed a significant reduction in variance of the scalar operator for the quenched Wilson-Dirac matrix. In our tests, we only examine the variance of the trace estimator and do not perform the full trace estimation. For performing the full trace estimation, see Ref. \cite{morgan23} on our new Multi-polynomial Monte Carlo method.

\section{Noise Subtraction Methods}




Measurements of operators $\Theta$ on the lattice come in the form of vacuum expectation values of said operators. These vacuum expectation values can be related to a trace calculation as, 
\begin{equation}
    \langle \bar{\psi} \Theta \psi \rangle = -Tr(\Theta M^{-1}), 
\end{equation}\noindent
where the Wilson-Dirac operator is given by $M = I - \kappa P$, $\kappa$ is the hopping parameter, and $P$ the hopping matrix. 
Due to the expense of inverting the quark matrix, this trace must be determined stochastically through Monte Carlo trace estimation. If one projects the quark matrix onto a set of $N$ noise vectors $\eta^{(n)}$ as, 
\begin{equation}
    Mx^{(n)} = \eta^{(n)},
\end{equation}\noindent
then the trace can be determined stochastically \cite{hutchinson89} by, 
\begin{equation}
    Tr(\Theta M^{-1}) = \frac{1}{N} \sum_{n=1}^{N} \eta^{(n) \dagger} \Theta x^{(n)},
\end{equation}
where, 
\begin{equation}
    x^{(n)} = M^{-1} \eta^{(n)}. 
\end{equation}
The purpose of noise subtraction methods is to reduce the variance of such trace estimators. With noise subtraction, we may relate the variance of the trace estimator to the off-diagonal elements of $M^{-1}$. Suppose we define a matrix $X$ made from $Z(N\geq 3)$ noise vectors $\eta^{(n)}$ such that, 
\begin{equation}
    X_{ij} \equiv \frac{1}{N} \sum^N_{n=1} \eta^{(n)}_i \eta^{*(n)}_j.   
\end{equation}\noindent
It can be shown \cite{baral19} that this matrix, along with a matrix representation of an operator $Q$, will produce the result, 
\begin{equation} \label{eq:var}
    V[Tr(QX_{Z(N \geq 3)})] = \frac{1}{N} \sum_{i \neq j} |q_{ij}|^2.
\end{equation}
In other words, the variance of $Tr(QX_{Z(N \geq 3)})$ only depends on the off-diagonal elements of $Q$.

With noise subtraction methods, we take advantage of the properties of the trace by introducing a traceless matrix $\tilde{Q}$ so that,s
\begin{equation}
    \langle Tr(QX) \rangle = \Big \langle Tr \Big \{ (Q - \tilde{Q}) X \Big \} \Big \rangle .
\end{equation}\noindent
Combining this with Eq. \ref{eq:var} we see that,
\begin{equation}  \label{eq:var2}
    V\Big [ Tr \Big \{ (Q - \tilde{Q}) X_{Z(N \geq 3)} \Big \} \Big ] = \frac{1}{N} \sum_{i \neq j} (|q_{ij} - \tilde{q}_{ij} |^2).
\end{equation}
In other words, if one constructs a traceless approximation of the inverted Wilson-Dirac matrix, $\tilde{M}^{-1}$, such that its off-diagonal elements are similar to $M^{-1}$, then the variance of the operator $\Theta$ can be reduced by subtracting $\tilde{M}^{-1}$ from ${M}^{-1}$ in the trace estimator. 

Thus, for some approximate matrix inverse $\tilde{M}^{-1}$, the trace estimator becomes, 
\begin{equation} \label{eq:trace_simple}
    \begin{aligned}
    Tr(\Theta M^{-1}) = & \frac{1}{N}\sum_n^N(
    \eta^{(n)\dagger}\Theta [x^{(n)}  - \tilde{x}^{(n)} ]) + Tr(\Theta\tilde{M}^{-1}),
    \end{aligned}
\end{equation}\noindent
where $x^{(n)} = M^{-1}\eta^{(n)}$, and $\tilde{x}^{(n)} = \tilde{M}^{-1} \eta^{(n)}$. The second term in Eq. \ref{eq:trace_simple} is a trace correction term, as the approximation $\tilde{M}^{-1}$ is not exactly traceless. This forms the basis for all of our subtraction methods, each of which involve a different approximation for the inverse of the Wilson-Dirac matrix $\tilde{M}^{-1}$. It can be additionally shown \cite{thron97} that for Z(4) noise specifically, the variance of the trace estimator is minimized. For this reason, although this result holds for Z($N\geq3$) noise vectors, we restrict ourselves to Z(4) noise, specifically.  

We summarize our methods below, which are discussed in more detail in \cite{baral19}. Our methods for noise subtraction include perturbative subtraction (PS) \cite{bernardson93} \cite{wilcox00}, eigenvalue subtraction (ES)\cite{guerrero}, polynomial subtraction (POLY)\cite{liu}, and hermitian-forced eigenvalue subtraction (HFES)\cite{guerrero_dis} or, equivalently, singular value deflation \cite{gambhir17} \cite{romero20}. While each produces modest improvements in the reduction of the variance, it's the combination of these methods that produces the greatest improvements. Namely, the hermitian-forced eigenvalue subtraction variants of PS and POLY which we refer to as HFPS and HFPOLY, respectively. 
For the HFPOLY method, the trace estimator $Tr(\Theta M^{-1})$ for an operator $\Theta$ and quark matrix $M$ can be \cite{baral19} expressed as, 
\begin{equation}
    \begin{aligned}
    Tr(\Theta M^{-1}) = & \frac{1}{N}\sum_n^N(
    \eta^{(n)\dagger}\Theta [x^{(n)}  - \tilde{x}'^{(n)}_{eig} - (\tilde{x}^{(n)}_{poly} - \tilde{x}'^{(n)}_{eigpoly}) ])
    + Tr(\Theta\gamma_5\tilde{M}'^{-1}_{eig})\\ 
    & + Tr(\Theta\tilde{M}'^{-1}_{poly} - \Theta\gamma_5\tilde{M}'^{-1}_{eigpoly}),
    \end{aligned}
\end{equation}\noindent
where, 
\begin{gather}
        x^{(n)} = M^{-1} \eta^{(n)} \\
        \tilde{x}^{(n)} \equiv \tilde{M}'^{-1}_{poly}\eta^{(n)}\\
        \tilde{x}'^{(n)}_{eig} \equiv \gamma_5\tilde{M}'^{-1}_{eig}\eta^{(n)}
        = \gamma_5\sum_q^Q \frac{1}{\lambda'(q)}e'^{(q)}_R(e'^{(q)\dagger}_R\eta^{(n)})\\
        \tilde{x}'^{(n)}_{eigpoly} \equiv \gamma_5\tilde{M}'^{-1}_{eigpoly}\eta^{(n)}
        = \gamma_5\sum_q^Q \frac{1}{\xi'(q)}e'^{(q)}_R(e'^{(q)\dagger}_R\eta^{(n)}).
\end{gather}\noindent
The vectors $e'^{(q)}_R$ are the eigenvectors of $M' \equiv M\gamma_5$, and $\lambda'(q)$ and $\xi'(q)$ are the eigenvalues of $\tilde{M}'_{eig}$ and $\tilde{M}'_{eigpoly}$, respectively. 

Furthermore, the approximate inverses to the quark matrix $\tilde{M}'^{-1}_{eig}$, $\tilde{M}_{poly}^{-1}$, $\tilde{M}'^{-1}_{eigpoly}$ 
are given by, 
\begin{gather}
    \qquad \tilde{M}^{-1}_{poly} \equiv b_0 I + b_1 \kappa P + b_2(\kappa P)^2 + b_3(\kappa  P)^3 + b_4(\kappa P)^4 + b_5(\kappa P)^5 + \ldots +  b_n(\kappa P)^n \\
    \tilde{M}'^{-1}_{eig} \equiv \tilde{V}'_R \tilde{\Lambda}'^{-1} \tilde{V}'^{\dagger}_R\\
    \tilde{M}'^{-1}_{eigpoly} \equiv \tilde{V}'_R \tilde{\Xi}'^{-1} \tilde{V}'^{\dagger}_R,
\end{gather}\noindent
where $P$ is the hopping matrix, $\kappa$ is the hopping parameter, $\tilde{V}'$ is a matrix whose columns are the $Q$ smallest right eigenvalues of $M'$, and $\tilde{\Lambda}'^{-1}$ and $\tilde{\Xi}'^{-1}$ 
are each diagonal matrices with $1/\lambda'^{(q)}$ and $1/\xi'^{(q)}$ along their diagonals, respectively. For the HFPS method, one replaces the approximate inverse $\tilde{M}^{-1}_{poly}$ with $\tilde{M}^{-1}_{pert}$. Further details about the methods can be found in Ref. \cite{baral19}. 

As shown in Eq. \ref{eq:var2}, the variance of the trace estimator does not depend on the trace correction term, $Tr(\Theta\tilde{M}'^{-1}_{poly} - \Theta\gamma_5\tilde{M}'^{-1}_{eigpoly})$. For this reason, while the term can be safely neglected when examining the effects of the method on variance reduction, we have separately \cite{morgan23} investigated the computation of this term which is necessary for the full trace estimator using our new Multipolynomial Monte Carlo approach. The scope of the current work is limited to the study of the variance reduction performance of the high-degree polynomials. In the following section, we discuss the implementation of the improved algorithm for forming and applying the subtraction polynomial beyond the $d = 7$ polynomials previously used in Ref. \cite{liu}. 

\section{The GMRES Polynomial}

The noise subtraction methods that use polynomials, POLY and HFPOLY, each involve forming an approximate inverse to the quark matrix as,  
\begin{equation}
    \qquad \tilde{M}^{-1}_{poly} \equiv b_0 I + b_1 \kappa P + b_2(\kappa P)^2 + b_3(\kappa  P)^3 + b_4(\kappa P)^4 + b_5(\kappa P)^5 + \ldots + b_n (\kappa P)^n,
\end{equation}
where the coefficients $b_0, b_1, \ldots, b_n$ are determined by the desired type of polynomial. 



For the choice of polynomial, we first construct the GMRES polynomial. The GMRES polynomial is, by definition, determined by the residual norm of the GMRES algorithm \cite{saad}. For iterative solvers of systems of linear equations, the residual norm is defined as, 
\begin{equation}
    ||r||_2 = ||b - M\hat{x}||, 
\end{equation}
where $\hat{x}$ is the approximate solution obtained from the solver. Krylov methods restrict the solution space of the solver to a Krylov subspace, which is to say, $\hat{x} \in \mathcal{K}_m(M,b) = span\{b, Mb, \ldots, M^{m - 1}b \}$. Thus, the iterative solution can be expressed in terms of a polynomial of the system, $\hat{x} = p(M) b$. From this, the residual norm becomes $||r||_2 = ||b - Mp(M)b|| = ||\pi(M)||$. In the GMRES algorithm, this polynomial $\pi(M)$, known as the GMRES polynomial, is the polynomial that minimizes the residual norm, and is defined by, 
\begin{equation}
    ||r||_2 = \min_{\pi \in \mathcal{P}} ||\pi(M)b||_2.
\end{equation}
Note that the GMRES polynomial $\pi(M) = I - Mp(M)$ minimizes the residual norm when $Mp(M) := I$. This behavior occurs when the polynomial $p(M)$ forms a good approximation for the inverse of $M$ such that the residual norm is minimized. This polynomial $p(M)$ is the polynomial previously \cite{liu} used in our noise subtraction methods for low-degree polynomials. In other words, $\tilde{M}^{-1}_{poly} = p(M)$.

The previous method used \cite{liu} for forming the GMRES polynomial involved making use of the Krylov matrix, 
\begin{equation}
    Y = \{b, Mb, M^2b, ..., M^m b \},
\end{equation}
and from that, forming the normal equations, 
\begin{equation}
    (MY)^T MYg = (MY)^T b,
\end{equation}
where the elements of the solution vector $g$ form the coefficients of the polynomial. This method of finding the polynomial, however, is unstable as the system becomes nearly singular for high-degree polynomials. 
Due to this, our previous subtraction results for POLY and HFPOLY \cite{baral}\cite{baral19}\cite{liu} were limited to degree $d = 7$. Our tests of POLY and HFPOLY on dynamical configurations were also limited to degree $d = 7$ \cite{whyte19}. 

We now implement a new algorithm \cite{morgan19} for forming $p(\alpha)$ which was originally used \cite{embree21} in the context of polynomial preconditioning. Because a polynomial preconditioner $p(M)$ also aims to approximate $M^{-1}$, we have chosen to use this same method to form $\tilde{M}^{-1}_{poly}$ as an alternative to using the normal equations. This was done through implementing the GMRES polynomial in the following factored form, 
\begin{equation}
    \pi(\alpha) = \prod_{i = 1}^{d} \Big( 1 - \frac{\alpha}{\theta_i} \Big), 
\end{equation}
where $\theta_1, \theta_2, \ldots, \theta_d$ are the Leja ordered \cite{reichel}, harmonic Ritz values obtained from a single cycle of GMRES($d$), and $d$ is the desired degree of the GMRES polynomial. This method of finding the polynomial is much more stable, allowing for high-degree polynomials to be formed. The GMRES polynomial $\pi(\alpha)$ is related to the subtraction polynomial through the connection, $\pi(\alpha) = 1 - \alpha p(\alpha)$. Algorithm 3 in Ref. \cite{embree21} provides details for applying the polynomial $p(\alpha)$ once the $d$ harmonic Ritz values have been obtained. Furthermore, from its connection with the GMRES polynomial, the polynomial to be used in noise subtraction $p(\alpha)$ will be of degree $d-1$ for a GMRES polynomial $\pi(\alpha)$ formed from $d$ Leja ordered, harmonic Ritz values. For this reason, we instead perform a single cycle of GMRES($d+1$) to obtain $d+1$ Leja ordered, harmonic Ritz values and, from these, form a degree $d$ noise subtraction polynomial $p(M)$ to approximate $M^{-1}$. Note that, in this work, we employ Leja ordering and not the Modified Leja ordering as discussed in Ref. \cite{embree21}. This is because the Leja points in Modified Leja ordering are assumed to be exact complex conjugate pairs, but the harmonic Ritz values of the Wilson-Dirac matrix will not come in exact complex conjugate pairs in floating-point arithmetic. 

\subsection{Reducing Costs Through Double Polynomials}
The degree of the subtraction polynomials are dependent on the size of the Krylov subpace used to form them. Since the GMRES algorithm is based on the Arnoldi iteration, forming large subspaces to produce high-degree subtraction polynomials can increase the orthogonalization and memory costs considerably. This is particularly true for larger lattices, where increasingly high-degree polynomials are needed. As a cost-saving measure, we may make use of double polynomials to avoid the large orthogonalization and memory costs of the Arnoldi iteration. To begin, we first outline double polynomials in the context of polynomial preconditioning, as they were originally \cite{embree21} used, and then we state how they can be used in noise subtraction. 

In polynomial preconditioning, we use a polynomial preconditioner $p_{in}(M)$ on the system $M$, yielding the polynomial preconditioned system $\phi_{in}(M) \equiv M p_{in}(M)$ as, 
\begin{equation}
     \phi_{in}(M) y = b,  
\end{equation}
where $x = p_{in}(M) y$. 
If we then polynomial precondition the system a second time using the preconditioner $p_{out}(\phi_{in}(M))$, we will have a double polynomial preconditioned system,
\begin{equation}
     \phi_{out}(\phi_{in}(M)) z \equiv \phi_{in}(M) p_{out}(\phi_{in}(M))z = b,  
\end{equation}
where $x = p_{in}(M) p_{out}(\phi_{in}(M))z$. 

If we consider $\phi_{out}(\phi_{in}(M)) = M p_{in}(M) p_{out}(\phi_{in}(M))$, we see that performing double polynomial preconditioning can also be thought of as applying a single polynomial preconditioner $p_{double}(M) \equiv p_{in}(M) p_{out}(\phi_{int}(M))$ to the original system $M$. 

To form such a double polynomial, we perform a single cycle of GMRES($d_{in}+1$) on $M$ to obtain $d_{in}+1$ harmonic Ritz values for $M$, and then perform a single cycle of GMRES($d_{out}+1$) on $\phi_{in}(M)$ to obtain $d_{out}+1$ harmonic Ritz values for $\phi_{in}(M)$. Note that the latter single cycle of GMRES($d_{out} + 1$) on $\phi_{in}(M)$ can alternatively be thought of as performing a single cycle of polynomial preconditioned GMRES or PP-GMRES($d_{out}+1$) on $M$ to form the $d_{out}+1$ harmonic Ritz values. From these, one can form polynomials $p_{in}(M)$ and $p_{out}(\phi_{in}(M))$ of degrees $d_{in}$ and $d_{out}$, respectively, using the algorithm discussed in Ref. \cite{embree21} to obtain a double polynomial of degree $d_{double} = (d_{in} + 1)*(d_{out} + 1) - 1$ for a preconditioner. In this way, we may form two low-degree polynomials $p_{in}(M)$ and $p_{out}(\phi_{in}(M))$ using smaller Krylov subspaces to achieve an effective high-degree polynomial, thus avoiding the large Krylov subspaces formed by the Arnoldi iteration of GMRES. 

In the context of variance reduction, we make use of this same double polynomial as our approximate inverse to the quark matrix $\tilde{M}^{-1}_{poly} \equiv p_{double}(M)$ to achieve high-degree polynomials at a lower orthogonalization and memory costs. We will compare the variance reduction effectiveness of this double polynomial to that of a single subtraction polynomial $p(M)$ of the same effective polynomial degree.

\section{Results}
\subsection{Measuring the Variance Reduction Performance}

In order to monitor the reduction in variance of each method over that of no subtraction (NS), we define the relative variance as, 
\begin{equation}
    \bar{\sigma}_{RE}^2 \equiv \frac{\bar{\sigma}_{A}^2}{\bar{\sigma}_{NS}^2},
    \label{eq:relvar}
\end{equation}
where $\bar{\sigma}_A^2$ is the variance obtained from a particular subtraction method averaged over the configurations, and $\bar{\sigma}^2_{NS}$ is the variance of the trace estimator with no subtraction averaged over all configurations.  

As we will see, the variance of the high-degree polynomial methods fluctuates across configurations as the polynomial degree is increased. For this reason, we also define a base-10 log-averaged relative variance using the geometric mean as, 
\begin{equation}
    \bar{\sigma}_{RE,log}^2 \equiv 10^{\bar{\sigma}^2_{A,log} - \bar{\sigma}^2_{NS,log}},
    \label{eq:logaverelvar}
\end{equation}
where $\sigma_{A,log}^2 = \frac{1}{N_{cfg}}\sum_{j=1}^{N_{cfg}} \log_{10}[(\sigma_{A}^2)^{(j)}]$, $(\sigma_{A}^2)^{(j)}$ is the variance obtained for configuration $j$ for a particular method, and $N_{cfg}$ is the number of configurations. In doing so, we aim to avoid a single configuration dominating the average by instead averaging the base-10 log of each variance estimate and subsequently re-exponentiating the result.

\subsection{Relative Variance and Deflation}

We begin by examining the variance reduction performance of high-degree polynomials when combined with deflation. To do so, we measure the scalar operator $Re[\bar{\psi}(x) \psi(x)]$ on a quenched lattice of size $24^3 \times 32$ consisting of 10 configurations run at $\beta = 6.0$ and with a hopping parameter of $\kappa = 0.1570 \approx \kappa_{crit}$. The Monte Carlo estimation of the variance used a total of $N = 100$, unpartitioned Z(4) noise vectors for each configuration. Previously \cite{baral}, our combo methods HFPOLY and HFPS performed the best, with HFES being the next best in error reduction. For this reason, we compare the methods NS, HFES, POLY, and HFPOLY. 

\begin{figure}[h!]
    \centering
    \begin{subfigure}[b]{0.45\textwidth}
    \includegraphics[trim={10cm 0 0 0},scale=0.07]{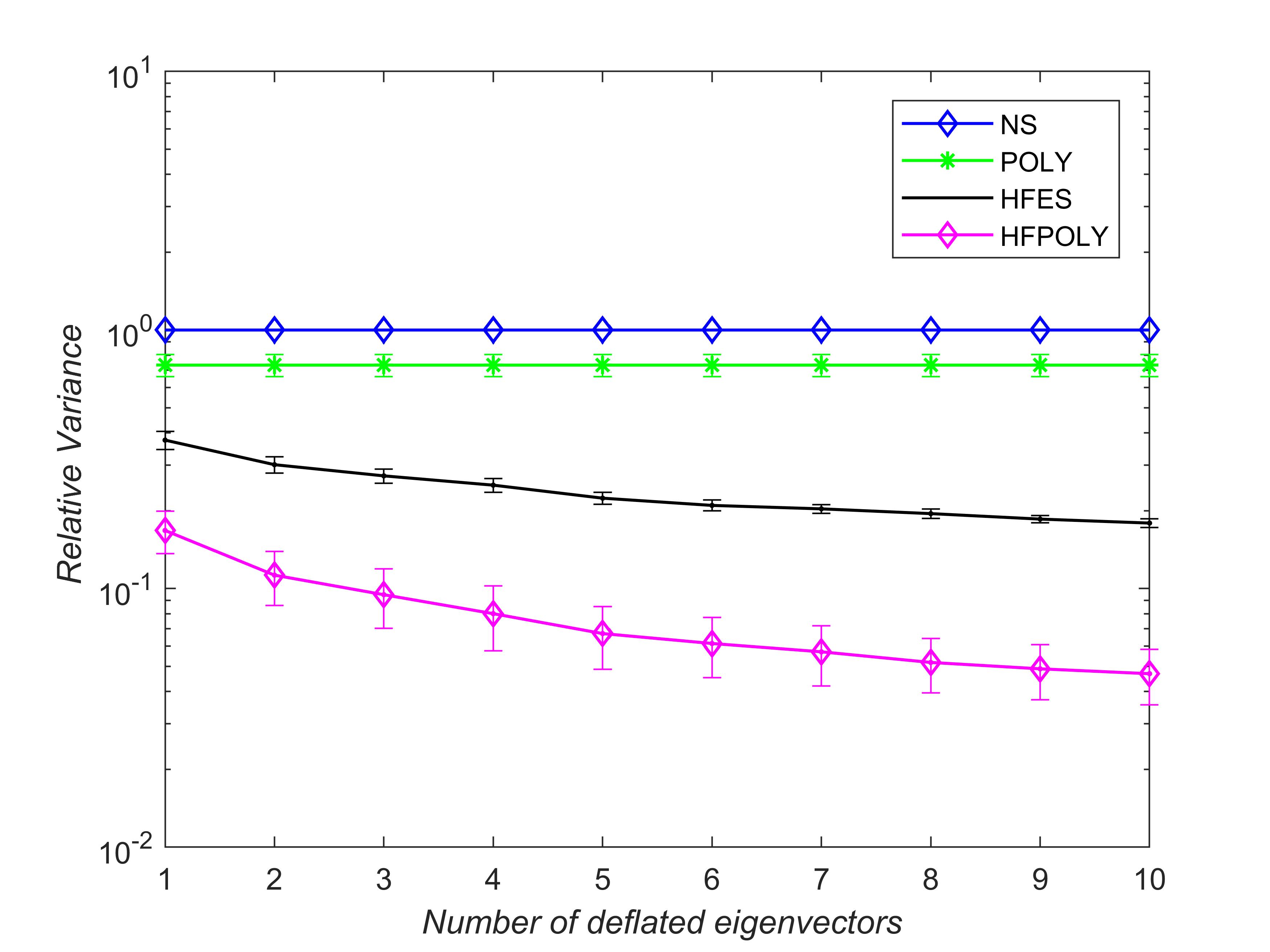}
    \caption{} 
    \label{fig:fig_1}
    \end{subfigure} 
    \hfill 
    \begin{subfigure}[b]{0.45\textwidth}
    \includegraphics[trim={15cm 0 0 0},scale=0.07]{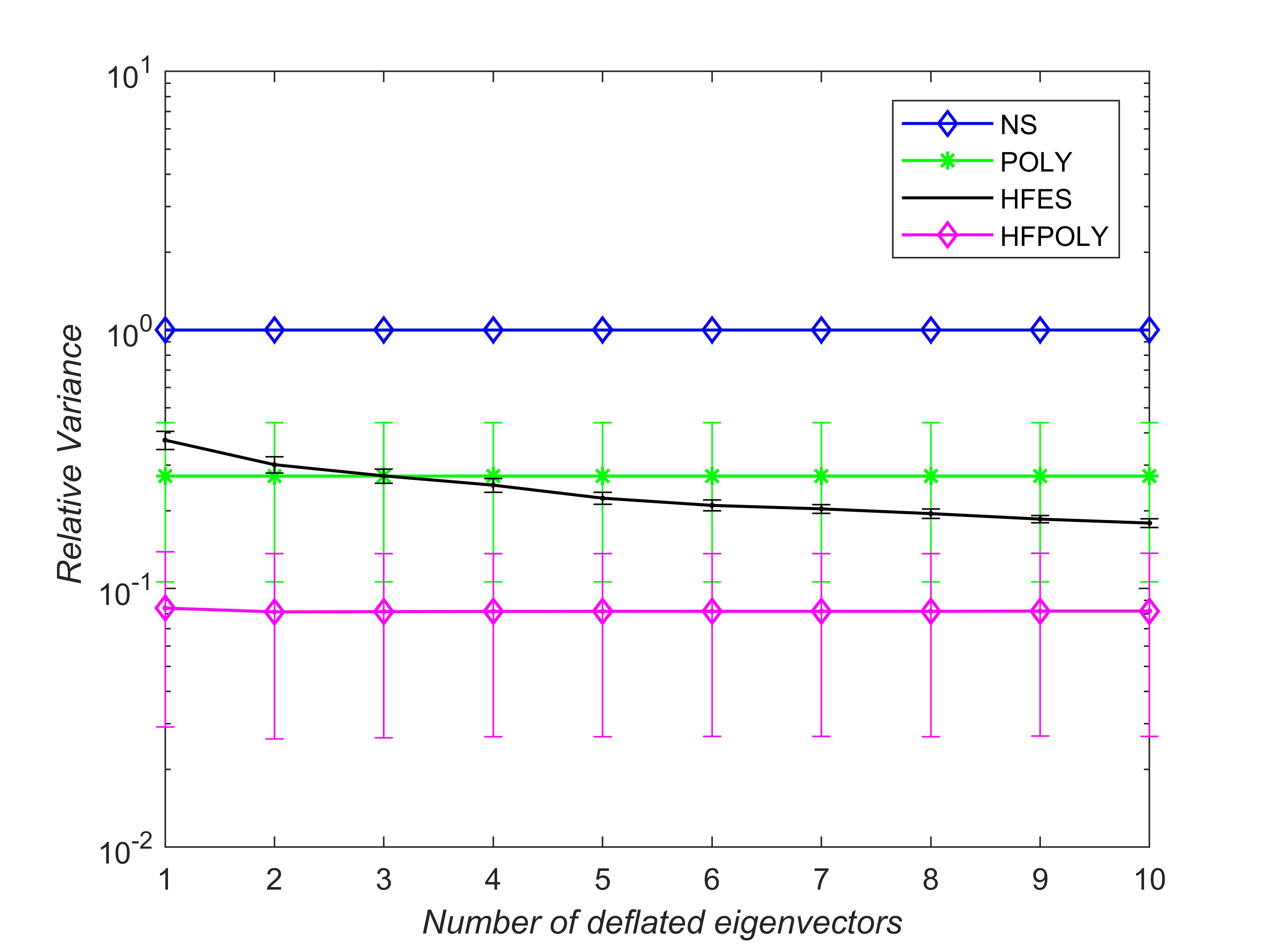}
    \caption{} 
    \label{fig:fig_2} 
    \end{subfigure} 
    \caption{Relative variance of the scalar operator $Re[\bar{\psi}(x)\psi(x)]$ as a function of deflated eigenvalues. Results averaged over 10 configurations of a $24^3 \times 32$ lattice at $\beta = 6.0$ and $\kappa = 0.1570$ with 100 noise vectors used in the Monte Carlo. Plots (a) and (b) correspond to degree $d = 200$ and $d = 1000$ subtraction polynomials, respectively.}
    \label{fig:deflation}
\end{figure}

Figure \ref{fig:deflation} shows the relative variance of the scalar operator versus the number of deflated eigenmodes for subtraction using polynomial degrees $d = 200$ and $d = 1000$, which were the smallest and largest degree polynomials examined for this lattice, respectively. Error bars correspond to the jackknife estimate of the error. Both POLY and HFPOLY push the variance of the scalar operator well below that of no subtraction as the polynomial degree is increased. Additional improvements through deflation, however, appear negligible past a single deflated eigenvalue when the polynomial degree is increased sufficiently high, beyond which deflation has little effect on variance reduction. In addition, the POLY method becomes more competitive against the HFES method as the polynomial degree is increased. Note, however, the significant increase in the error bars of POLY and HFPOLY as the polynomial degree is increased. Furthermore, in (b) of Figure \ref{fig:deflation}, the relative variance of HFPOLY after deflating $k = 10$ eigenmodes appears to slightly increase once it has reached a high enough polynomial degree compared to (a) in Figure \ref{fig:deflation}. This difference, however, is still within error. Another point of interest is that, since the deflation appears to no longer have a significant effect on the variance reduction of HFPOLY, it appears that HFPOLY will approach the same variance results as POLY for sufficiently high polynomial degrees, and that the improvements of HFPOLY over POLY will become even less significant. We will return to this point shortly when we examine the relative variance as a function of the polynomial degree.

\subsection{Dependence on Polynomial Degree and Lattice Volume}
We now turn our attention to considering the polynomial degree dependence of our POLY and HFPOLY noise subtraction methods. To begin, we first examined the relative variance for POLY which was determined for several polynomial degrees on several lattice volumes. Here, we once again consider the scalar operator but with three additional quenched lattices of volumes $4^3 \times 4$, $8^3 \times 8$, and $12^3 \times 16$ with each run at $\beta = 6.0$ at $\kappa = 0.1570 \approx \kappa_{crit}$ using a total of 10 configurations for each lattice volume. The three smaller lattices used $N = 10$, unpartitioned Z(4) noise vectors for the Monte Carlo while the largest used $N = 100$.

\begin{figure}[h!]
    \centering
    \includegraphics[scale=0.7]{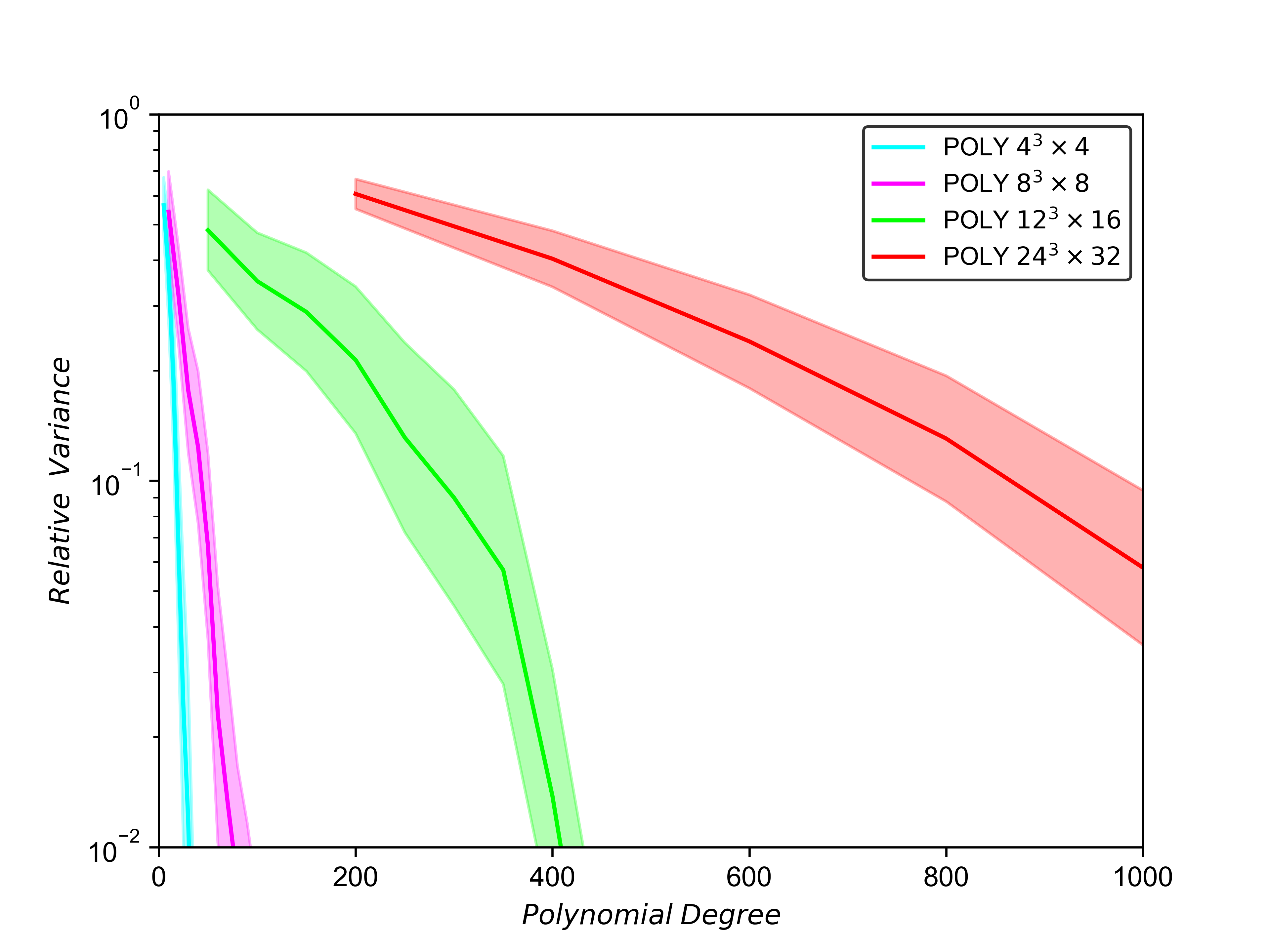}
    \caption{Log-averaged Relative variance of the scalar operator using the POLY method against the degree of the subtraction polynomial. Lattice volumes $4^3 \times 4$, $8^3 \times 8$, $12^3 \times 32$, and $24^3 \times 32$ shown, each averaged over 10 quenched configurations at $\beta = 6.0$ and $\kappa = 0.1570$.}
    \label{fig:re_var_deg_allvols}
\end{figure}

We observed a large variation in the relative variance across configurations when using a fixed polynomial degree for each of the configurations and so, for this reason, we choose to use log-averaging for the relative variance as defined in Eq. \ref{eq:logaverelvar} so that one configuration does not dominate the average.
The jackknife error of the log-averaged relative variance that we obtained, however, is still negative and, as such, cannot be plotted on a log plot. Instead, our measurement is of the exponent of $\bar{\sigma}^2_{RE,log}$ or simply, $\langle log_{10}[\bar{\sigma}^2_{RE,log}] \rangle = log_{10}[\bar{\sigma}^2_{RE,log}] \pm \delta(log_{10}[\bar{\sigma}^2_{RE,log}])$, where the uncertainty comes from performing the jackknife. When plotting, we re-exponentiate $\bar{\sigma}^2_{RE,log}$ to obtain the log-averaged relative variance. For the error bars, we instead plot for the length of the top error bars, 
\begin{equation} 
T = 10^{(\bar{\sigma}^2_{RE,log} + \delta(\bar{\sigma}^2_{RE,log}))} - 10^{(\bar{\sigma}^2_{RE,log})},
\label{eq:top}
\end{equation}
and for the length of the bottom error bars, 
\begin{equation}
B = 10^{(\bar{\sigma}^2_{RE,log})} - 10^{(\bar{\sigma}^2_{RE,log} - \delta(\bar{\sigma}^2_{RE,log}))}.
\label{eq:bottom}
\end{equation} 
Figure \ref{fig:re_var_deg_allvols} shows the log-averaged relative variance of the scalar operator versus the polynomial degree for each of the four lattice volumes that we obtained, where the solid line corresponds to the log-averaged relative variance and the shaded error bars correspond to the error obtained using Eqs. \ref{eq:top} and \ref{eq:bottom}. 
We found that POLY reduced the variance of the trace estimator by an order of magnitude compared to no subtraction. As seen in Figure \ref{fig:deflation}, HFPOLY is also capable of reducing the variance by an order of magnitude compared to no subtraction using a degree $d = 1000$ polynomial for the largest lattice when using Eq. \ref{eq:relvar} as opposed to Eq. \ref{eq:logaverelvar}.


As can be seen, the increase in volume requires a higher polynomial degree to achieve the same variance reduction performance. This polynomial degree can become impractically high for larger lattices. To increase the polynomial degree further, one would need to reduce the expense of computing such high-degree polynomials, as the single cycle of GMRES($d+1$) involved in forming the subtraction polynomial of degree $d+1$ becomes expensive due to the orthogonalization costs of the Arnoldi iteration. As will be seen shortly, double polynomials as discussed in Ref. \cite{embree21} \cite{morgan19} can be used to combat this.

\begin{figure}[h!]
    \centering
    \includegraphics[width=.73\textwidth]{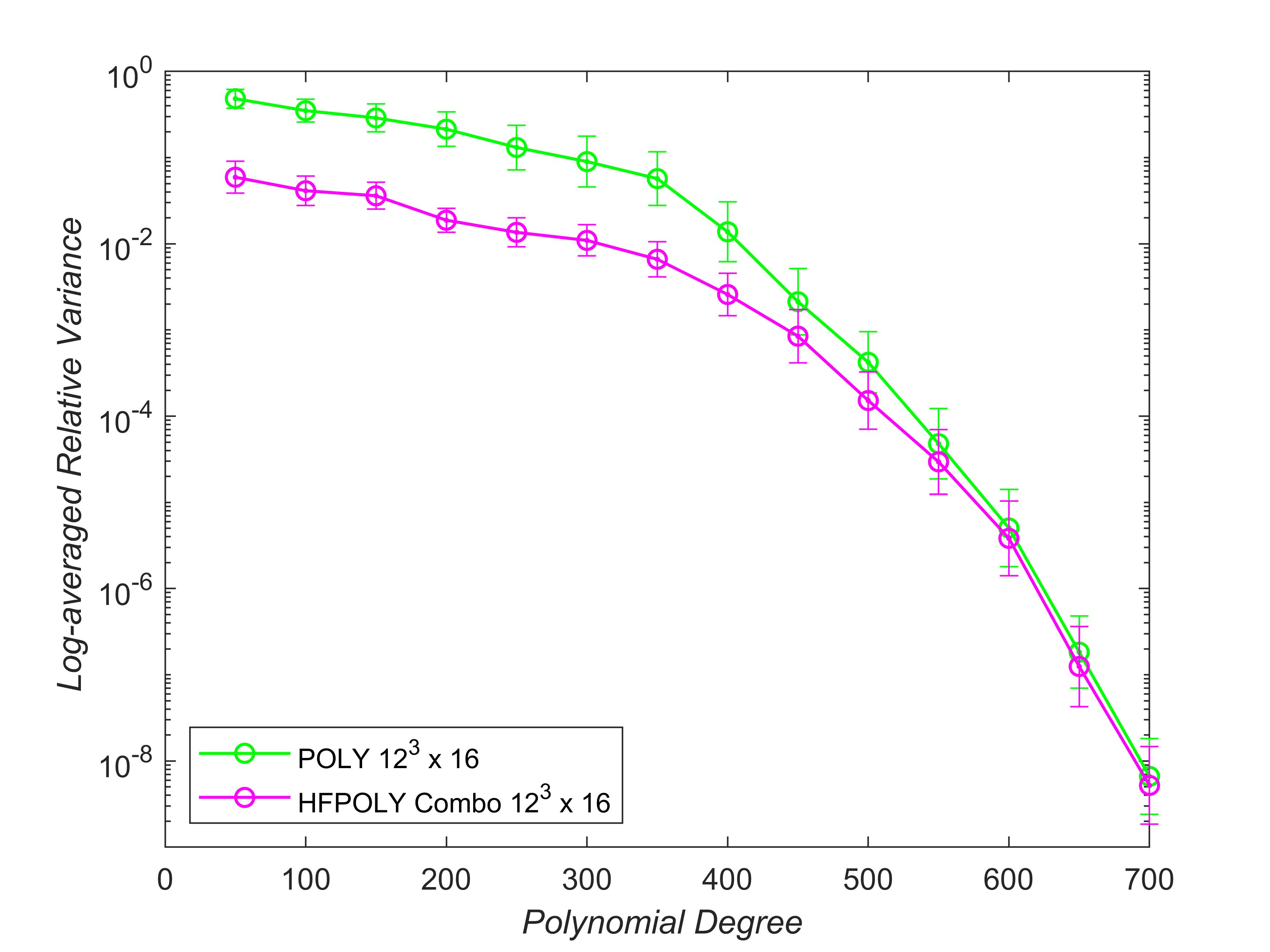}
    \caption{Log-averaged relative variance for POLY and HFPOLY methods as a function of polynomial degree at $\kappa_{crit}$ for the scalar operator: $Re[\bar{\psi}(x)\psi(x)]$. Lattice volume for both methods is $12^3 \times 16$ and results were averaged over 10 configurations using 10 noise vectors. A total of $k = 5$ eigenmodes are deflated for the HFPOLY combo method. Error bars determined according to Eq. \ref{eq:top} and \ref{eq:bottom}.}
    \label{fig:re_var_12121216}
\end{figure}

In addition, we examined the dependence on the polynomial degree for both the POLY and HFPOLY methods for the $12^3 \times 16$ lattice shown in Figure \ref{fig:re_var_12121216}. Here, we also use the same log-averaged relative variance for the solid line, and the error bars once again come from Eqs. \ref{eq:top} and \ref{eq:bottom}. Note that the gap between the two methods appear to disappear as the polynomial degree is increased. For sufficiently high polynomial degrees, the improvements of deflation to HFPOLY over POLY become less significant as the polynomial degree becomes high enough to effectively approximate the inverse to the original system. As mentioned before, similar behavior to this is seen in Figure \ref{fig:deflation} in the comparison between the degree 200 and 1000 relative variance reduction plots for the $24^3 \times 32$ lattice. There, we saw that as the polynomial degree increased, the gap between POLY and HFPOLY begins to narrow. We anticipate that for sufficiently high-degree polynomials, the relative variance results for the POLY and HFPOLY on the $24^3 \times 32$ lattice will become nearly identical, as well. To examine such high-degree polynomials for the $24^3 \times 32$ lattice, however, would require using double polynomials, as single polynomials of such high-degree are impractical.

\subsection{Dependence on GMRES Relative Residual Norm}

The large variation across configurations seen for the relative variance of a given polynomial degree suggests that selecting the same polynomial degree to use for every configuration may not be the most effective means of forming the polynomial. Rather than fix the polynomial degree for each configuration, we may instead let the polynomial degree become variable across configurations by monitoring the relative residual norm of the single cycle of GMRES used to form the polynomial. 

For each configuration, GMRES was run for a single cycle and the relative residual norm for the linear equations was monitored until the subspace reached a size large enough that the relative residual norm reached a desired tolerance level. The resulting size of the Krylov subspace determined the degree of the GMRES polynomial, and the harmonic Ritz values were stored to be used in forming the polynomial. This produces a variable polynomial degree for each configuration, but with each having reached the same relative residual norm from GMRES. In doing so, rather than examining the relative variance as a function of the polynomial degree, we may instead determine the relative variance as a function of the relative residual norm tolerance used during the single cycle of GMRES when constructing the polynomial. 

\begin{figure}[h!]
    \centering
    \includegraphics[scale=0.7]{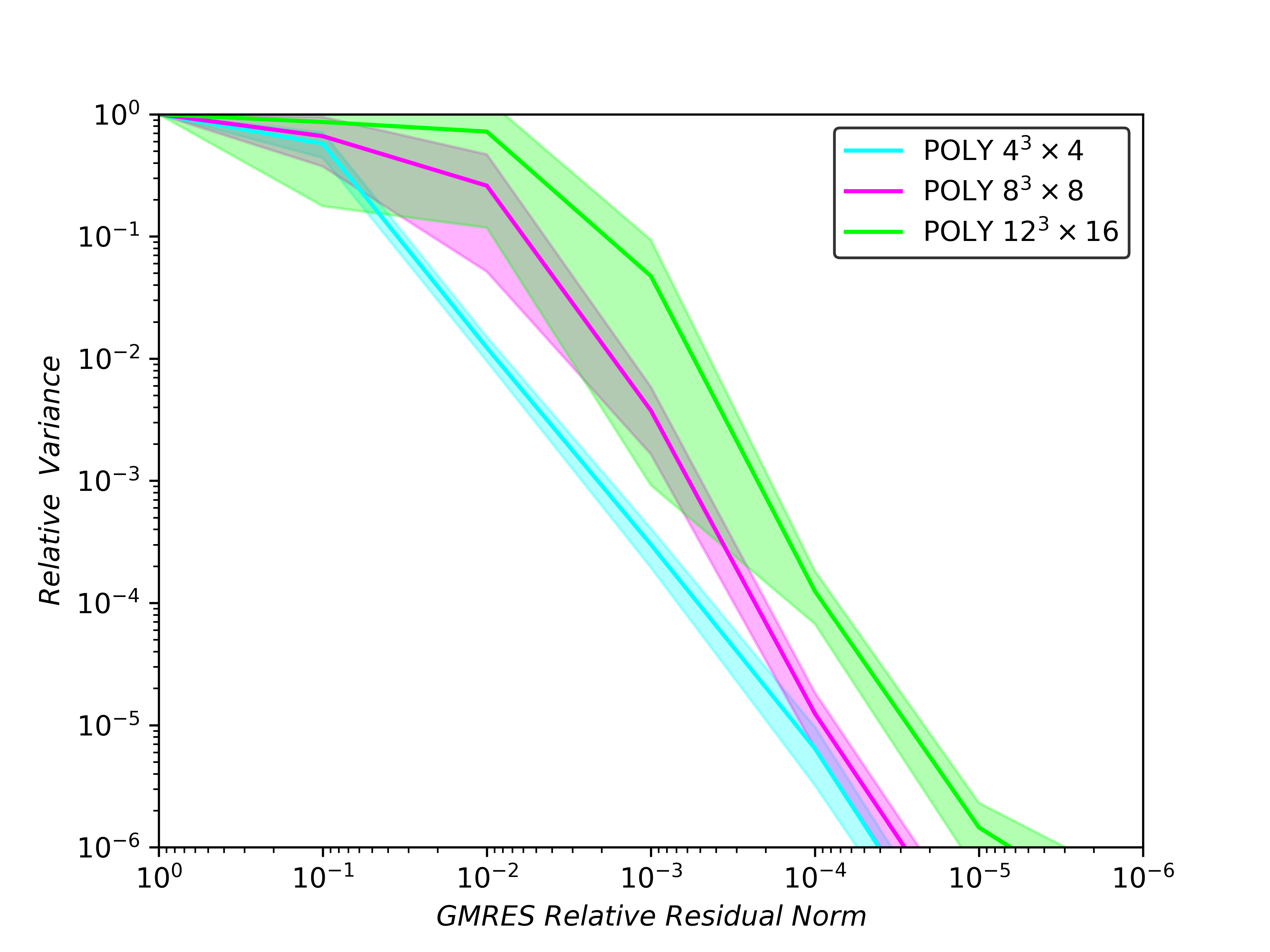}
    \caption{Relative variance of the scalar operator using the POLY method as a function of the GMRES relative residual norm tolerance used to form the subtraction polynomial. The three quenched lattice volumes $4^3 \times 4$, $8^3 \times 8$, $12^3 \times 16$ are shown, each at $\beta = 6.0$ and $\kappa = 0.1570$. A total of $N = 10$ noise vectors used in the Monte Carlo}
    \label{fig:re_var_rn_allvols}
\end{figure}

In Figure \ref{fig:re_var_rn_allvols}, we show the relative variance obtained from the three smallest lattice volumes as a function of the relative residual norm achieved from running a single cycle of GMRES. Here we are determining the relative variance using Eq. \ref{eq:relvar} without the use of log-averaging. The shaded regions correspond to a simple jackknife error across the configurations.  

We see significantly less variation across configurations when the GMRES relative residual norm is used to specify the polynomial degree, rather than choosing the same polynomial degree for each configuration. This is especially true once the GMRES relative residual norm has been pushed to a sufficiently low tolerance. This suggests that setting a tolerance for the GMRES relative residual norm instead of explicitly specifying the same polynomial degree or subspace size for all of the configurations provides a better method for selecting the polynomial degrees. We make use of this method of monitoring the GMRES relative residual norm in our work on the Multipolynomial Monte Carlo method \cite{morgan23}, as it is important for how we automate the generation of the subtraction polynomial for each configuration in that work. 

\subsection{Variance Reduction With Double Polynomials}
\begin{figure}[h!]
    \centering
    \includegraphics[width=.73\textwidth]{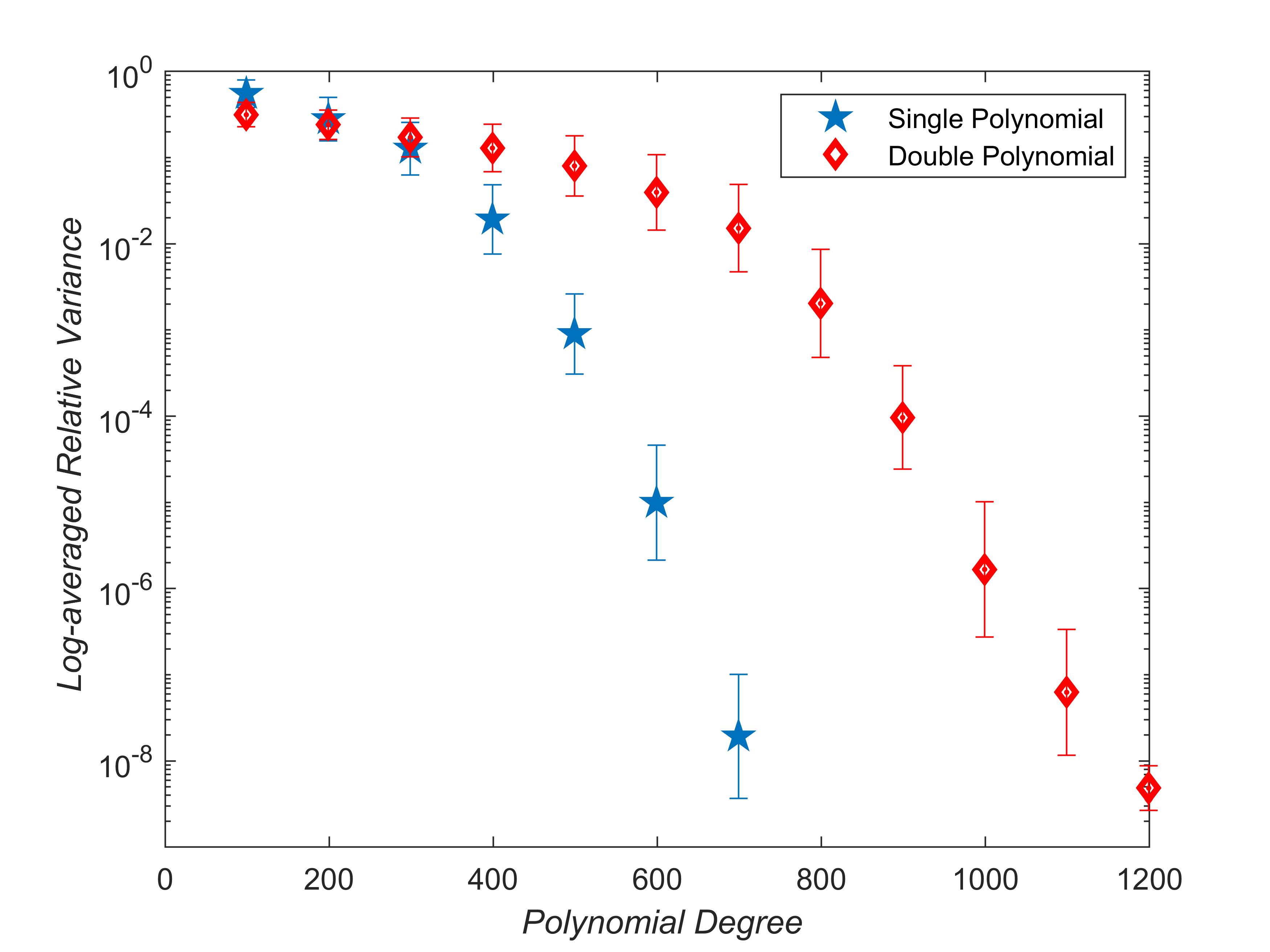}
    \caption{Log-averaged relative variance for the scalar operator using a single polynomial and double polynomial as a function of polynomial degree. Lattice volume for both methods is $12^3 \times 16$ on the 10 quenched configurations at $\beta = 6.0$ and $\kappa = 0.1570$. A total of $N = 10$ noise vectors were used in the Monte Carlo.}
    \label{fig:doublepolydemo}
\end{figure}

As a final test, we seek to improve on the high-degree polynomial method by using double polynomials as a cost-saving measure, which we compare against a single polynomial. In Figure \ref{fig:doublepolydemo}, we show the log-averaged relative variance of the scalar operator as a function of the polynomial degree using ten configurations of a $12^3 \times 16$ lattice at $\kappa = 0.1570 \approx \kappa_{crit}$ and $\beta = 6.0$. Error bars correspond to those outlined in Eq. \ref{eq:top} and Eq. \ref{eq:bottom}. The Monte Carlo is performed using $N = 10$, unpartitioned Z(4) noise vectors, while the noise vector used to form the polynomial is instead a normalized Gaussian noise vector. For the double polynomial, we choose to set the inner polynomial $\phi_{in}(M)$ to degree $d_{in} + 1 = 20$. The outer polynomial degree $d_{out}$ is chosen such that $d_{double} = [99, 199, 299, 399, 499, 599, 699, 799, 899, 999, 1099, 1199]$. For the single polynomial, we use the degrees $d = [99, 199, 299, 399, 499, 599, 699]$ for comparison.

As can be seen, at lower polynomial degrees both single and double polynomials produce similar variance reduction results, but as the degree increases, the difference between them also increases. The double polynomial requires a significantly higher polynomial degree to achieve the same variance reduction performance as a single polynomial, but this high-degree polynomial is much cheaper to form, and such cost-saving measures will be necessary for larger lattices where forming a single high-degree polynomial is impractical. This demonstrates one means of improving the cost-effectiveness of such high-degree polynomials in noise subtraction, and is a vital component to our Multipolynomial Monte Carlo method \cite{morgan23}.

\section{Conclusions}
With the use of high-degree polynomials, the previously obtained noise reduction for our POLY and HFPOLY methods are greatly improved. While both methods reduced the variance below that of no subtraction as the polynomial degree increased, deflation became less effective beyond a single deflated eigenvalue. It was also observed that the POLY and HFPOLY methods appear to approach to the same noise reduction performance as the polynomial degree is increased. The relative variance for the scalar operator was only investigated up to degree $d = 1000$ polynomials for the $24^3 \times 32$ lattice. A large amount of fluctuation across configurations was observed when fixing the polynomial degree for each configuration. This variation was reduced by, instead, using a variable polynomial degree across configurations by monitoring the GMRES relative residual norm from the single cycle as a stopping criteria for selecting the polynomial degree. To increase the polynomial degree beyond this point, we have also demonstrated the potential for using double polynomial methods to reduce the expense of the Arnoldi iteration in the single cycle of GMRES required to form the polynomial. These double polynomials will help reduce MPI communication and storage costs associated with orthogonalization of Krylov subspace basis vectors in GMRES at the cost of reducing the variance reduction effectiveness of polynomial noise subtraction. For large lattices, however, such cost-reduction measures are necessary and are a vital component to our new Multipolynomial Monte Carlo method \cite{morgan23}.

\section{Acknowledgements}
Numerical work was performed using the High-Performance Computing (HPC) Cluster at Baylor University. We also acknowledge the Texas Advanced Computing Center (TACC) at The University of Texas at Austin for providing HPC resources that have contributed to the research results reported within this paper. URL: http://www.tacc.utexas.edu. We would like to thank Dr. Yue Ling for graciously allowing us time on his personal compute nodes. We also would like to thank Abdou Abdel-Rehim, Victor Guerrero, Suman Baral, and Travis Whyte for their contributions to this work. We also thank Randy Lewis for his QQCD program.


\begin{thebibliography}{99}


\bibitem{reichel} 
Z. Bai, D. Hu, and L. Reichel, 
IMA J. Numer. Anal. \textbf{14}, 563 (1994). 


\bibitem{baral19}
S. Baral, T. Whyte, W. Wilcox, and R. Morgan,
Comput. Phys. Comm. \textbf{241}, 64 (2019).


\bibitem{baral} 
S. Baral, W. Wilcox and R. B. Morgan,
arXiv:1611.01536 [hep-lat].


\bibitem{bernardson93} 
S. Bernardson, P. McCarty, and C. Thron, 
Comput. Phys. Commun. \textbf{78}, 256 (1993).


\bibitem{morgan08}
D. Darnell, R. B. Morgan, and W. Wilcox,
Lin. Alg. and its Appl. \textbf{429}, 2415 (2008).

\bibitem{embree21}
M. Embree, J. A. Loe, and R. B. Morgan, 
SIAM J. Sci. Comput. \textbf{43}, A1 (2021). 

\bibitem{hutchinson89}
M.  F.  Hutchinson, 
Commun. Stat. Simul. Comput. \textbf{18}, 1059 (1989).

\bibitem{gambhir17}
A. S. Gambhir, A. Stathopoulos, and K. Orginos, SIAM J.
Sci. Comput. \textbf{39}, A532 (2017).


\bibitem{guerrero}
V. Guerrero, R. B. Morgan, W. Wilcox, 
``Eigenspectrum noise subtraction methods in lattice QCD", \textit{Proceedings}, \textit{The XXVII International Symposium on Lattice Field Theory (Lattice 2009): Beijing, China, July 26-31, 2009},
PoS \textbf{LATTICE2009}, 041 (2010), arXiv:1001.4366 [hep-lat].


\bibitem{guerrero_dis}
V. Guerrero, \textit{Electric Neutron Polarizability and Eigenspectrum Subtraction Techniques for Disconnected Quark Loops}, PhD Dissertation, (Baylor University, Waco, 2011). 


\bibitem{liu}
Q. Liu, W. Wilcox, and R. B. Morgan, 
arXiv:1405.1763 [hep-lat].


\bibitem{liupp}
Q. Liu, R. B. Morgan, W. Wilcox, SIAM J. Sci. Comput. \textbf{37}, S407 (2015). 


\bibitem{morgan19}
J. A. Loe and R. B. Morgan, 
Numer. Linear Algebra Appl. \textbf{29}, 1 (2021). 



\bibitem{morgan02}
R. B. Morgan, 
SIAM J. Sci. Comput. \textbf{24}, 20 (2002). 


\bibitem{morgan23}
R. B. Morgan, W. Wilcox, P. Lashomb, T. Whyte, (2023).


\bibitem{romero20}
E. Romero, A. Stathopoulos, and K. Orginos, 
J. Comput. Phys. \textbf{409}, 109356 (2020).



\bibitem{saad} 
Y. Saad and M. H. Schultz, 
SIAM J. Sci. Stat. Comput. \textbf{7}, 856 (1986).


\bibitem{thron98}
C. Thron, S. Dong, K. F. Liu, and H. P. Yeng, 
Phys. Rev. D 
\textbf{57}, 1642 (1998). 


\bibitem{thron97}
C. Thron, K. F. Liu, S. J. Dong, 
Nucl. Phys. B (Proc. Suppl.) \textbf{53}, 977 (1997). 


\bibitem{whyte19}
T. Whyte, S. Baral, P. Lashomb, W. Wilcox, R. B. Morgan,
arXiv:1912.02878 [hep-lat]. 


\bibitem{wilcox00}
W. Wilcox, 
Nucl. Phys. B (Proc. Suppl.) 
\textbf{83}, 834 (2000).  


\end{thebibliography}
\end{document}